\documentstyle[aps,prl]{revtex}
\begin{document}
\draft
\title{Post freeze-out annihilations in the early universe}
\author{Steen Hannestad}
\address{Theoretical Astrophysics Center, Institute of Physics and 
Astronomy,
University of Aarhus, 
DK-8000 \AA rhus C, Denmark}
\date{\today}
\maketitle

\begin{abstract}
We investigate the possible
effect of the residual annihilations of heavy particles 
after freeze-out from equilibrium in the early universe. 
An error in previous work on this subject is pointed
out and the correct method of solving the Boltzmann equation
for this case is developed. For Majorana particles there are significant
differences relative to previous work. 
\\keywords: Early universe - dark matter, nucleosynthesis
\end{abstract}

{\it Introduction:}
In the early universe particles are subject to both pair-annihilation
and elastic scattering interactions. At sufficiently early times
these interactions are usually strong enough to keep the particles
in complete thermodynamic equilibrium.
However, as the universe expands and cools, interactions become weaker
and eventually any particle species will decouple from equilibrium.
A quick estimate of the freeze-out epoch can be found by comparing the 
expansion timescale of the universe, $H^{-1}$, 
with the interaction timescale for
the given species, $\Gamma^{-1}$ \cite{kolb}. 
If the interaction timescale become 
longer than the expansion timescale the species decouples so that the
freeze-out criterion is
\begin{equation}
\frac{\Gamma}{H} \simeq 1.
\end{equation}
For massless particles with standard weak interactions one finds for
instance a freeze-out temperature of $T_F \simeq 1$ MeV.
If the species is absolutely stable against decay, its present
contribution to the energy density of the universe can basically
be found from the ratio $m/T_F$ (except for the possible dilution
of the species by entropy production subsequent to freeze-out).
This can be used to constrain models with stable massive particles
by demanding that their present day energy density does not overclose the
universe \cite{kolb}.
The standard example of this is the Weakly Interacting Massive Particle
(WIMP), a hypothetical heavy particle species 
(for instance the lightest supersymmetric particle) which could make up
the dark matter of our universe.

Even if a given particle species violates this constraint it may still
be allowed if it decays on a sufficiently short timescale.
There are, however, other possible ways of constraining
such scenarios. For example, a massive decaying neutrino
 will in general change the outcome of Big Bang nucleosynthesis (BBN)
\cite{DGT94}.
Further, if its decay products have electromagnetic content the model can 
be ruled out by other observations \cite{R96}.

Now, even for a stable particle there may be other possible constraints
than the energy density argument.
After freeze-out there continues to be some residual annihilations
of these particles. If their annihilation
products interact electromagnetically and are sufficiently energetic
they may photodisintegrate light elements
 and ruin the agreement between BBN theory and 
observations. For this reason BBN can be used to constrain the 
interaction strength and mass of WIMPs.
This has been done several times in the literature, both for high
mass particles ($m \gtrsim 1 {\rm GeV}$) \cite{RS88,RPH88,HP90},
and for more moderate masses \cite{FKT90,KR96}.
In all cases, the so-called integrated Boltzmann equation has been used
to determine the annihilation rate of heavy particles. This equation 
assumes that all particles are in kinetic equilibrium,
but, as will
be pointed out, it has not been taken properly into account that any
massive particle species which decouples from chemical equilibrium 
eventually also decouples from kinetic equilibrium.
Using the standard integrated Boltzmann equation thus leads to wrong
estimates of the annihilation rate.

{\it The integrated Boltzmann equation:}
The approach to take when calculating the freeze-out of massive
species is to use the Boltzmann equation \cite{kolb,bernstein}.
\begin{equation}
\partial f = C_{\rm ann}+C_{\rm el},
\label{eq:boltz}
\end{equation}
where
\begin{equation}
\partial f = \frac{\partial f}{\partial t} - H p \frac{\partial f}
{\partial p},
\end{equation}
 $H$ being the Hubble parameter $H \equiv \dot{R}/R$.
On the right-hand side,
 $C_{\rm ann}$ and $C_{\rm el}$ represent annihilations and elastic
scatterings respectively.
If one assumes that 
the interaction is $CP$ conserving
these collision terms can be written generically as \cite{kolb}
\begin{eqnarray}
C_{\text{coll}}[f] & = & \frac{1}{2E_{1}}\int d^{3}\tilde{p}_{2}
d^{3}\tilde{p}_{3}d^{3}\tilde{p}_{4}
\Lambda(f_{1},f_{2},f_{3},f_{4}) \nonumber \\ 
& & \,\,\, \times S \sum | \! M \! |^{2}_{12\rightarrow 34}\delta^{4}
({\it p}_{1}+{\it p}_{2}-{\it p}_{3}-{\it p}_{4})(2\pi)^{4}, 
\label{integral}
\end{eqnarray}
as long as we are only concerned with 2-body collisions of the type
 $1+2 \to 3+4$.
Here we have $d^{3}\tilde{p}
 = d^{3}p/((2 \pi)^{3} 2 E)$. $S$ is a symmetrisation
factor of 1/2! for each pair of identical particles in initial or 
final states \cite{wagoner},
and $\sum \mid\!M\!\mid^{2}$ is the interaction matrix element
squared and spin-summed.
 ${\it p}_{i}$ is the four-momentum of particle $i$ and 
 $\Lambda(f_{1},f_{2},f_{3},f_{4}) = f_3 f_4(1\pm f_1)(1 \pm f_2) -
f_1 f_2(1\pm f_3)(1 \pm f_4)$ is the phase-space factor with 
 $+$ corresponding to Fermi-Dirac statistics and $-$ to 
Bose-Einstein statistics.

If one then assumes scattering equilibrium for all particles involved
the elastic scattering terms all equate to zero.
Further, a significant simplification is possible if one uses Boltzmann
statistics for all particles instead of the appropriate quantum
statistics, so that
\begin{equation}
1 \pm f \to 1,
\end{equation}
where, again, + corresponds to Fermi-Dirac statistics and $-$ to 
Bose-Einstein statistics.
This reduces the phase-space factor to
\begin{equation}
\Lambda(f_{1},f_{2},f_{3},f_{4}) = f_3 f_4 - f_1 f_2
\end{equation}
The last assumption is that the annihilation products are in full
thermodynamic equilibrium so that
\begin{equation}
f_3 f_4 = e^{-(E_3+E_4)/T} = e^{-E_1/T}e^{-E_2/T} = 
f_{1,{\rm eq}} f_{2,{\rm eq}}.
\end{equation}
By integrating over momentum space for the incoming particle one then
arrives at the well known integrated Boltzmann equation 
\cite{kolb,bernstein}
\begin{equation}
\dot{n} = - 3 H n - \langle \sigma v \rangle (n^2 - n_{\rm eq}^2),
\label{eq:intboltz}
\end{equation}
where $\langle \sigma v \rangle$ is the velocity-averaged annihilation
cross-section, defined as \cite{kolb,GG91}
\begin{eqnarray}
\langle \sigma v \rangle & = & \frac{1}{n_{\rm eq}^2}
\int d^{3}\tilde{p}_{1} d^{3}\tilde{p}_{2}
d^{3}\tilde{p}_{3}d^{3}\tilde{p}_{4}
f_{1,{\rm eq}}f_{2,{\rm eq}} \nonumber \\ 
& & \,\,\, \times S \sum | \! M \! |^{2}_{12\rightarrow 34}\delta^{4}
({\it p}_{1}+{\it p}_{2}-{\it p}_{3}-{\it p}_{4})(2\pi)^{4},
\label{eq:cross}
\end{eqnarray}
and $n_{\rm eq}$ is the equilibrium number density
\begin{equation}
n_{\rm eq} = \frac{g}{2 \pi^2} \int_0^\infty p^2 f_{\rm eq}(p) dp.
\end{equation}
Below we list the assumptions going into the integrated Boltzmann 
equation,
Eq.~(\ref{eq:intboltz}): \\
1: The heavy particle species is in scattering equilibrium \\
2: All annihilation products are in thermodynamic equilibrium \\
3: Boltzmann statistics instead of Fermi-Dirac or Bose-Einstein 
statistics \\
4: $CP$ invariance. \\

At very high temperatures the heavy species will be in complete
thermal equilibrium. However, as $T$ drops below the rest mass $m$,
the rate for pair annihilation and production becomes exponentially
suppressed, $\Gamma \propto e^{-m/T}$, quickly leading to freeze-out
of chemical equilibrium at some temperature, $T_{F_A}$.

However, the rate for scattering on other relativistic species is
{\it not} exponentially suppressed. This means that kinetic equilibration
is maintained long after freeze-out of chemical equilibrium, but
at some point there will be a freeze-out of scattering equilibrium
too. From that point on the distribution will remain fixed in comoving 
momentum space
\begin{equation}
f(p,T) = f(p_F T_{F_S}/T,T_{F_S}),
\end{equation}
where $T_{F_S}$ indicates the temperature for freeze-out of scattering
equilibrium.
At the point where scattering freezes out the distribution has the
form
\begin{equation}
f(p,T_{F_S}) = e^{-(m+p^2/2m-\mu)/T_{F_S}}.
\end{equation}
At later times, this distribution is no longer of equilibrium
shape. Rather it is an equilibrium distribution corresponding to
a new temperature parameter
\begin{equation}
T_* = T_F \frac{T^2}{T_F^2}.
\end{equation}
Since the integrated Boltzmann equation only applies to species in 
kinetic
equilibrium it is not apparent that it should be applicable to 
a treatment of post freeze-out annihilation. 
Note, however, that for a non-relativistic species after freeze-out,
 $f_{\rm eq} \ll f$, so that pair-production can be neglected. Thus, the 
annihilation term in the full Boltzmann equation reads
\begin{eqnarray}
C_{\text{ann}}[f] & = & - \frac{1}{2E_{1}}\int d^{3}\tilde{p}_{2}
d^{3}\tilde{p}_{3}d^{3}\tilde{p}_{4}
(f_1 f_2) \nonumber \\ 
& & \,\,\, \times S \sum | \! M \! |^{2}_{12\rightarrow 34}\delta^{4}
({\it p}_{1}+{\it p}_{2}-{\it p}_{3}-{\it p}_{4})(2\pi)^{4}, 
\end{eqnarray}
where $f$ is a distribution function corresponding to scattering 
equilibrium with a temperature $T_*$.
Thus, one can still use the integrated Boltzmann equation in the shape
\begin{equation}
\dot{n} = - 3 H n - \langle \sigma v \rangle n^2,
\end{equation}
but with a different definition of $\langle \sigma v \rangle$.
An annihilation cross section which in kinetic equilibrium is 
proportional
to $T^n$ becomes instead $\langle \sigma v \rangle \propto T_*^n$.
In general we can therefore state that after chemical freeze-out there
will be a temperature region, $T_{F_A} \geq T \geq T_{F_S}$, 
where the equilibrium assumption
 $\langle \sigma v \rangle \propto T^n$ applies. Then at $T=T_{F_S}$
kinetic equilibrium decouples and the thermally averaged cross
section follows the new relation $\langle \sigma v \rangle \propto 
T_*^n$.

All previous work has used the equilibrium assumption, 
 $\langle \sigma v \rangle \propto T^n$, even after kinetic
freeze-out, which gives a significantly different temperature dependence
of the annihilation rate.
For Dirac particles, $n=0$, and 
the result is the equivalent to the standard
one. For Majorana particles,$n=1$
\footnote{This applies as long as the massive species is non-relativistic
and the annihilation products are massless.},
and there is now a big difference
between $\langle \sigma v \rangle \propto T$ and
 $\langle \sigma v \rangle \propto T_*$. The cross section falls off much
faster with decreasing temperature so that residual annihilations are
weaker than found in the standard calculation.

How big this effect is depends on the difference between
the kinetic and chemical equilibration rates.
In general scattering and pair processes arise from the same
coupling, so that the fundamental interaction strength is the same
for the two different processes.
The thermally averaged scattering cross section will therefore be
 $\langle \sigma v \rangle \propto m T$.
However, what we are really after is the thermalisation rate for the
heavy particle, not its interaction rate. In general a non-relativistic
particle scattering on a massless species only gains or looses 
a fraction $T/m$
of its energy. Therefore the real thermalisation cross section should
instead be
 $\langle \sigma v \rangle \propto T^2$.
Since the relativistic species have number densities going as $T^3$
the overall thermalisation rate can be estimated as
\begin{equation}
\Gamma_T \propto T^5.
\end{equation}
This rate is in general much larger than the annihilation rate because
it does not have an exponential suppression factor in the number density
of scatterers. The ratio of annihilation to thermalisation can
be written approximately as
\begin{equation}
\frac{\Gamma_A}{\Gamma_T} \simeq \left(\frac{m}{T}\right)^{\frac{7}{2}-n}
e^{-m/T},
\end{equation}
as long as scattering equilibrium still holds.
For a Dirac particle decoupling from chemical equilibrium 
when $m/T \simeq
20$ this ratio is roughly $10^{-4}$ at chemical freeze-out.

{\it Applications:}
There are a number of cases where the above formalism may be applicable.
The standard example, of course, is the massive neutrino.
Frieman {\it et al.} \cite{FKT90} have developed an analytic 
estimate of the 
effect of residual annihilations
\footnote{Their treatment of the photon cascade processes is
not numerically accurate and lacks several relevant processes, as pointed
out in Ref.~\cite{subir}. However, since we are not after high precision
in our calculations we shall use the results of Ref.~\cite{FKT90} anyway.}.
Their results only apply
to masses below the threshold for neutron and proton production 
($m \lesssim 1$ GeV),
but will serve well to illustrate our results (for a treatment of higher
masses, see for instance Refs.~\cite{RPH88,HP90}).
Energetic particles injected into the cosmic plasma after BBN can
photodisintegrate the light elements. The primary effect is to fission
 $^4$He to produce $^3$He and D. Since the primordial element abundances 
are quite well determined one cannot, for instance,
tinker too much with the ratio of
helium isotopes without coming into conflict with observations.

The primordial value of $^4$He has been determined by
Olive and Steigman \cite{OS} to be
\begin{equation}
Y_P = 0.232 \pm 0.003 ({\rm stat}) \pm 0.005 ({\rm syst}),
\end{equation}
where $Y_P$ is the mass fraction of helium. Other determinations have been
slightly different, but since we are not interested in great accuracy
we shall just use the above value without further discussion.
For $^3$He the primordial value is bounded from below 
by observations of the local interstellar medium to be \cite{hata}
\begin{equation}
\frac{N(^3 {\rm He})}{N(H)} \leq \frac{N(^3 {\rm He + D})}{N(H)} 
\leq 1.1 \times 10^{-4}. 
\end{equation}
One then has an upper limit to the ratio $^3$He/$^4$He of
\begin{equation}
\frac{N(^3 {\rm He})}{N(^4 {\rm He})} \lesssim 2 \times 10^{-3} 
\label{eq:helim}
\end{equation}

As an example we take the annihilation of very massive neutrinos
with standard weak interactions. For both Dirac and Majorana neutrinos
one finds that for a mass of the order 1 GeV, the chemical freeze-out
happens at $x_F \simeq 17$ giving a chemical decoupling temperature,
 $T_{F_A}$, of 60 MeV. 
For Majorana neutrinos the scattering equilibrium decoupling
happens at $T_{F_S} \simeq 1$ MeV.

Using the constraint above on the helium isotope ratios 
Frieman {\it et al.} \cite{FKT90}
have derived an estimate of how the mass and interaction rate of a given
particle can be constrained
\begin{eqnarray}
m_{\rm MeV} & \gtrsim & 1.0 \times 10^{-2} \, e \, x_F^2 
\left[\frac{10^{-38}\, {\rm cm}^2}{\sigma_0}\right]^{1/(n+1)}\nonumber \\
&& \,\,\, 
\times \left( \frac{2.6 \times 10^4 B}{g_*(T_F) \Omega_B h^2 
(n+\frac{3}{2})(n+\frac{5}{2})}
\right)^{1/(n+1)}.
\label{eq:primhe}
\end{eqnarray}
Here, the following parametrisations have been used:
 $\langle \sigma v \rangle = \sigma_0 (T/m)^n$ and the number of massless 
degrees of freedom is defined as $g_* = \frac{30}{\pi^2} T^{-4} \rho_R$.
 $B$ is the branching ratio to electromagnetic annihilation products
and $x_F \equiv m/T_F$.

The above constraint was derived assuming that the equilibrium
version of the Boltzmann equation is correct.
However, it turns out that we can still use it for Majorana neutrinos.
The photo destruction of Helium only begins after $T=1$ MeV \cite{FKT90}
which means
that at the time of $^4$He
destruction the Majorana neutrinos have also decoupled
from scattering equilibrium. This means that 
 $\langle \sigma v \rangle \propto T^2$. However, we cannot just 
use the above relation taking $n=2$ because that would assume that
the scattering equilibrium decoupled at the same time as the
chemical equilibrium. 
If we first rescale the overall annihilation cross section with a factor
 $\psi \equiv T_{F_A}/T_{F_S} \simeq 60$ then we can use the constraint,
Eq.~(\ref{eq:primhe}), by using the substitutions
\begin{eqnarray}
\sigma_0 & \to & \sigma_0 x_F \psi, \\
n = 1 & \to & n = 2.
\end{eqnarray}

Plugging in numbers one finds that for a Majorana neutrino with standard
weak interactions there is a lower limit to the mass of
\begin{equation}
m_\nu \gtrsim 150 {\rm MeV}\,\, (n=2),
\end{equation}
which should be compared to the value one finds by using $n=1$
\begin{equation}
m_\nu \gtrsim 300 {\rm MeV}\,\, (n=1).
\end{equation}
For a Dirac neutrino the mass limit is of course unchanged and is found to
be \cite{FKT90}
\begin{equation}
m_\nu \gtrsim 1 {\rm GeV}\,\, (n=0).
\end{equation}
The above limits have been derived using $\Omega_B h^2 = 0.05$, $B=0.5$.

Thus, using the correct version of the Boltzmann equation weakens the
mass limit on heavy Majorana neutrinos by about a factor of two.
In itself, of course, the above limits are not that interesting because
massive neutrinos heavier than $m \simeq 20$ MeV are already ruled out
if they couple to the weak interaction with normal strength.
However, this weakening of the mass limit can be expected
to apply generally because it only depends on the different temperature
dependence of the annihilation cross section.

Other groups have studied the effect of WIMP annihilations by means of
numerical work \cite{RPH88,HP90}, 
taking into account for example the production of neutrons
and protons before or during nucleosynthesis. 
This approach, however, makes it very difficult to estimate the effect
of using the correct Boltzmann equation, though
in general the impact of WIMP annihilations will, as mentioned,
be less drastic than usually thought
because the cross-section decreases much faster with decreasing
temperature. This will make any bound on particle masses and/or
interaction strengths looser than has previously been found.

{\it Acknowledgements:}
I wish to thank Paolo Gondolo for interesting discussions on this subject.
In particular I wish to thank the referee for pointing out that assuming
that scattering and chemical equilibrium decouples at the same
temperature is wrong and that it leads to a substantial overestimate of
the importance of the effect described here.


\begin{references}
\bibitem{kolb} E.~W.~Kolb and M.~S.~Turner, {\it The Early Universe}, 
Addison Wesley (1990).
\bibitem{DGT94}S.~Dodelson, G.~Gyuk 
and M.~S.~Turner, Phys.\ Rev.\ D\ {\bf 49}, 5068 (1994);
M.~Kawasaki {\it et al.}, Nucl.\ Phys.\ {\bf B419},
105 (1994);
S.~Hannestad, Phys.\ Rev.\ D {\bf 57}, 2213 (1998).
\bibitem{R96}For a review of constraints on such decays, see for instance
G.~G.~Raffelt, {\it Stars as Laboratories for 
Fundamental Physics}, University of Chicago Press (1996).
\bibitem{RS88}M.~H.~Reno and D.~Seckel, Phys.\ Rev.\ D {\bf 37}, 
3441 (1988).
\bibitem{RPH88}J.~S.~Hagelin, R.~D.~J.~Parker and A.~Hankey, 
Phys.\ Lett.\ {\bf B215}, 397 (1988).
\bibitem{HP90}J.~S.~Hagelin and R.~D.~J.~Parker,
Nucl.\ Phys. {\bf B329}, 464 (1990).
\bibitem{FKT90}J.~A.~Frieman, E.~W.~Kolb and M.~S.~Turner,
Phys.\ Rev. D {\bf 41}, 3080 (1990).
\bibitem{KR96}E.~W.~Kolb and A.~Riotto,
Phys.\ Rev.\ D {\bf 54}, 3722 (1996). 
\bibitem{bernstein} J.~Bernstein, {\it Kinetic Theory in the Expanding 
Universe}, Cambridge University Press (1988).
\bibitem{wagoner}R.~V.~Wagoner, in {\it Physical Cosmology} - Les Houches 
(1979), edited by R. Balian, J. Audouze and D. N. Schramm,
\bibitem{GG91}P.~Gondolo and G.~Gelmini,
Nucl.\ Phys. {\bf B360}, 145 (1991); see also
J.~Edsj{\"o} and P.~Gondolo, Phys.\ Rev.\ D {\bf 56}, 1879 (1997).
\bibitem{subir}S.~Sarkar, Rep.\ Prog.\ Mod.\ Phys.\
{\bf 59}, 1493 (1996).
\bibitem{OS} K.~A.~Olive and G.~Steigman, 
Astrophys.\ J.\ Suppl.\ {\bf 97}, 49 (1997).
\bibitem{hata}N.~Hata {\it et al.}, Phys.\ Rev.\ Lett.\
{\bf 75}, 3977 (1995).
\bibitem{hannestad2}S.~Hannestad, submitted to Nucl.\ Phys. B (1998).
\end{references}
\end{document}